# Bell's Theorem and the Causal Arrow of Time

Nathan Argaman[*]

Physics department, N.R.C.N., P.O. Box 9001 Be'er Sheva 84190, ISRAEL.

(April 7, 2010)

Einstein held that the formalism of Quantum Mechanics (QM) entails "spooky actions at a distance." Indeed, in the 60's Bell showed that the predictions of QM disagree with the results of *any* locally causal description. Accepting non-local descriptions while retaining causality leads to a clash with the theory of relativity. Furthermore, the causal arrow of time by definition contradicts time-reversal symmetry. For these reasons, some authors (Feynman and Wheeler, Costa de Beauregard, Cramer, Price) have advocated abandoning microscopic causality. In the present article, a simplistic but concrete example of following this line of thought is presented, in the form of a retro-causal toy-model which is stochastic and which provides an appealing description of the specific quantum correlations discussed by Bell. One concludes that Einstein's "spooky actions" may occur "in the past" rather than "at a distance," resolving the tension between QM and relativity, and opening unexplored possibilities for future reformulations of QM.

---

[*] E-mail: argaman@mailaps.org .

"… nobody understands quantum mechanics."

R.P. Feynman[1]

I. Introduction

Quantum Mechanics (QM) has a remarkable number of mathematical formulations – Heisenberg's matrix mechanics,[2] Schrödinger's wave mechanics,[3] Feynman path integrals,[4] de Broglie-Bohm guiding waves,[5] Nelson's stochastic mechanics,[6] and more[7] – and many interpretations – the Copenhagen interpretation,[8] Everett's many worlds interpretation,[9] Cramer's transactional interpretation[10] and others.[11]  In this maze, the voice of John Bell speaks with exceptional clarity.[12]  In 1964, he provided a simple proof for

> Bell's theorem:[13]  The predictions of QM can not be reproduced by any locally causal mathematical description.

This is a major obstacle to understanding QM, in addition to other well-known difficulties such as the measurement problem,[14] and the exponential complexity of the QM description of many particles.  Naturally, much discussion of the possibility of hidden assumptions in Bell's analysis has ensued.[15]

Bell's work followed on a review of the hidden-variables problem in QM,[16] and is based on two-particle correlations,[17] generalizing those considered by Einstein, Podolsky and Rosen[18] (EPR) in 1935.  As a result, many of the subsequent publications have invoked assumptions that entered these earlier discussions, such as determinism[19] and realism.[20]  Such assumptions are not necessary for a proof of Bell's theorem.  However, there is one necessary assumption which is only implicitly stated in the bulk of the literature regarding Bell's theorem, including the original, and this is the assumption of causality,[21] i.e., the causal arrow of time.  In terms of a critique of the logic used, this is a very minor point, as the assumption of locality, which is made explicit, clearly presupposes causality.  Thus, the flaw in the analysis[22] may be adequately corrected by simply taking "locality" to be an abbreviation of "local causality."  Indeed, the latter term is the one preferred by Bell in later publications.[23]  However, the prevalent omission of an explicit mention of causality has consequences.

The problem becomes acute when special relativity and Lorentz invariance are brought into the discussion.  In fact, Bell himself concluded that it would be necessary to choose a preferred reference frame in which the non-local effects of QM would occur, defying the principles of relativity. For a while, Bell even advocated a return to the concept of the Aether.[24] The conclusion that a tension exists between the two pillars of 20th century physics, QM and the theory of relativity, is widely echoed to this day.[25] However, the tension is based on three ingredients – (a) QM, (b) relativity and (c) the causal arrow of time – and of these three, our confidence in the arrow of time should clearly be the weakest. After all, one is considering the microscopic realm, which is supposedly time-reversal symmetric...

The purpose of the present work is to review Bell's original 1964 contribution,[13] emphasizing rather than neglecting the assumption of causality (section II). As an "illustration" of the character of the correlations between distant particles predicted by QM, Bell included in his presentation a simplistic toy-model, which was necessarily non-local. In section III, this process of illustration is completed by introducing a similarly simplistic (but not as artificial) retro-causal toy-model, which reproduces the same predictions. Surprisingly, such simple and explicit retro-causal models are not to be found in the literature,[26,27,28] despite the facts that (a) restoring time-reversal symmetry at a fundamental level was achieved for classical charged particles by Wheeler and Feynman,[29] and for QM by Aharonov and coworkers,[30] and (b) the relevance of retro-causation to the EPR discussion was noted repeatedly since 1953 by Costa de Beauregard,[31,32,33] was made the basis of a re-interpretation of QM by Cramer[10] in the 1980s, and was thoroughly discussed by Price in the 1990s.[34] Naturally, these developments have been criticized by invoking interesting possibilities for causal loops.[35] Assuming that such loops can be avoided, it may be hoped that the idea of retro-causation may one day lead to yet another reformulation of QM, which could lead to significant progress in our understanding of QM, perhaps analogous to that associated with path integrals.[4] Further concluding remarks are provided in Section IV.

II. Bell's theorem

In order to briefly review Bell's theorem, we will describe a specific physical system, spell out the predictions QM makes for the correlations it exhibits, write down the most general expression for these correlations which could be generated by a locally causal mathematical model, and show that this expression is incompatible with the QM predictions. The topic has been reviewed many times, but the assumption of causality is generally not stated adequately. We will deviate from Ref. 13 by considering correlated photons, rather than spin one-half particles, as the expressions are slightly simpler (contain no minus signs), many of the experimental tests involved photons,[36] and photon polarization is more generally familiar than spin.

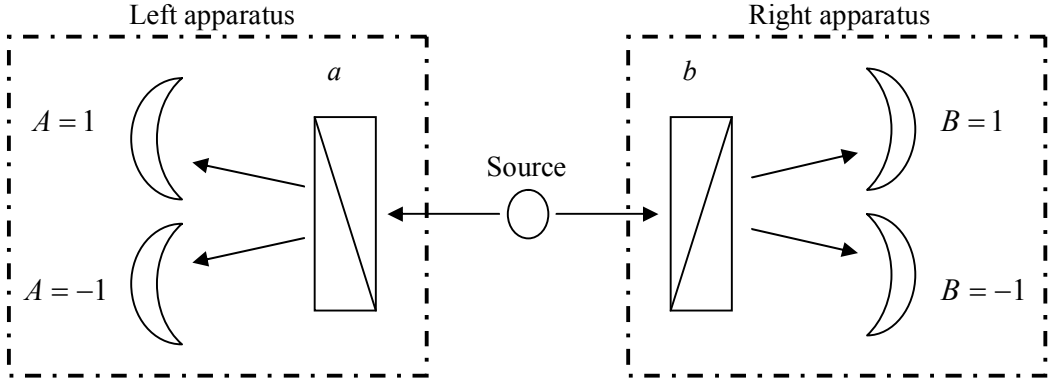

Fig 1: Schematic sketch of the setup considered by Bell, adapted to photons. The circle represents the source of entangled pairs of photons, the dot-dashed boxes represent the measuring apparatuses, and the arrows mark the paths of the photons. Each apparatus contains a polarizing beam splitter (PBS, crossed rectangle), and a pair of detectors (crescents). The angles $a$ and $b$ extend out of the plane of the figure, and measure the preferred polarization direction of each PBS (say, relative to the vertical). If the photon on the left (or the right) has the preferred polarization, it is deflected towards the detector labeled $A=1$ (or $B=1$), whereas if it has the perpendicular polarization, it reaches the other detector, resulting in $A=-1$ (or $B=-1$). The elements of the setup are considered perfect, so that shortly after each time the source is activated, two detectors click, one on the left and one on the right, and the associated values of $A$ and $B$ may be recorded. For large distances between the source and the apparatuses, the apparatus settings ($a$ and $b$) can be changed while the photons are in flight. The statistics of the results $A$ and $B$ are described by Eqs. (1) and (2), and can be measured by repeating the experiment many times.

Consider then an idealized experimental setup involving a source of pairs of photons and two spatially-separated measurement apparatuses, as sketched in Fig. 1. Every time the source is activated, two photons are emitted, with one reaching the left apparatus and the other reaching the right apparatus. The photon polarizations are described by a QM state $\psi$ of vanishing total angular momentum and even parity,[37] corresponding, e.g., to emission from Ca atoms, as used in many experiments.[36] These atoms are activated by excitation into a state of zero angular momentum, $J=0$, and emit the two photons through a radiative cascade, passing through an intermediate state with $J=1$ before reaching their $J=0$ ground state. Each measurement apparatus contains a polarizing beam splitter, with a preferred orientation denoted by $a$ for the one on the left, and by $b$ for the one on the right. These orientations are angles, defined modulo $\pi$. The measurement result on the left is denoted $A=1$ if the photon is detected to be polarized along $a$, and $A = -1$ if its polarization is found to be perpendicular to $a$. The result of the measurement on the right is similarly denoted by $B$.

For a given choice of the orientations $a$ and $b$ (the free variables), QM provides the probabilities, $P(A,B \mid a,b,\psi)$, for the four alternative experimental outcomes, $A = \pm 1, B = \pm 1$. It is convenient to specify these probabilities by stating the expectation values of the individual outcomes,

$$\langle A \rangle = \langle B \rangle = 0 \, , \tag{1}$$

and the correlator,

$$P_{\text{QM}}(a,b) \equiv \langle AB \rangle = \cos(2a - 2b) \, . \tag{2}$$

The QM description is slightly different if the measurement on one side, say, on the left, is performed before the other, although Eqs. (1) and (2) are retained. In such cases the probabilities on the left are evaluated through an expression of the form $P(A \mid a,\psi)$, and the wavefunction "collapses" into a simpler one, $\varphi(A,a,\psi)$, which is then used to evaluate the probabilities on the right, in the form $P(B \mid b,\varphi)$. Notice that even if the non-local "collapse" is avoided, as in the many-worlds interpretation, and $\psi$ is taken to evolve smoothly into a combination of $\varphi(+,a,\psi)$ and $\varphi(-,a,\psi)$, the description must still use an expression of the form $P(B \mid b,\varphi)$ in order to reproduce the predictions for the correlator, Eq. (2). It is thus non-local in the sense relevant to the present discussion – the evolution of $\psi$ may be "local" in configuration space, but it is not local in the 3-D physical space.[38] Of course, the non-locality of QM can not be used to send signals immediately from one location to another. This no-signaling condition is manifested here in Eq. (1), which specifies, e.g., that the probabilities for the different outcomes on the right do not depend on the free parameter on the left, $\langle B \rangle = 0$ regardless of $a$.

In order to discuss general mathematical descriptions conforming with local causality, one has to fully specify input and output variables, and possibly internal variables, and to associate space-time locations with each of them. Inputs are taken as causes and outputs as effects – in order to be able to identify the meaning of the causal arrow of time, one must avoid using a definition which automatically takes causes to precede their effects.[22] For example, Newton's equations with the initial positions and momenta treated as inputs form a causal description, as an external force applied at an intermediate time will have consequences only for later times, while this classical description becomes retro-causal if the *final* position and momenta are treated as free input variables instead. For the set-up considered, $a$ and $b$ are inputs, $A$ and $B$ (or rather, their probability distribution) are outputs, and they are all associated with the time of measurement. Bell introduced the notation $\lambda$ for the set of all the properties of each pair of photons just before the measurement is made on them. In general, this set may be empty, it may contain discrete or continuous variables, or it may also contain functions or more complicated constructs. In standard QM, one has a single possibility, $\lambda = \psi$, but a general description may involve stochastic elements, requiring a probability distribution $\rho(\lambda)$. Still more generally, one could consider a distribution of the type $\rho(\lambda \mid a, b)$, but this will be taken up only in the next section, as the assumption of *causality* would be violated if $\lambda$ (or its distribution) were to depend on the free variables associated with the time of measurement.[39] The assumption of *locality* implies that $A$ does not depend on the orientation $b$ of the distant beam-splitter or the measurement result $B$ there, and similarly, $B$ does not depend on $a$ and on $A$. Thus, a general locally causal description would specify, in addition to the character of $\lambda$, distributions $\rho(\lambda)$, $P(A|a,\lambda)$ and $P(B|b,\lambda)$ which must be non-negative and normalized. The correlator of interest may then be expressed as

(3) $$\langle AB \rangle = \int d\lambda\, \rho(\lambda) \sum_{AB} AB\, P(A|a,\lambda) P(B|b,\lambda) ,$$

where the integration over $\lambda$ is understood as a summation if $\lambda$ varies over a discrete set (and is absent if $\lambda$ has only one possible value).

Bell went through three steps to prove that the predictions of QM for the two-photon correlator, Eq. (2), can not be reproduced in this manner. The first step follows EPR in deducing that $A$ (and similarly, $B$) cannot be stochastic, and must instead be completely determined by $a$ and $\lambda$. Indeed, when $a = b$ QM predicts perfect correlations between the measurements of the two photons – the value of $A$ must in this case be equal to $B$ with a probability of 100%. Thus, when $b = a$, the probabilities $P(A|a,\lambda)$ in Eq. (3) must be equal to either 0% or 100%, and one may use a function $A(a,\lambda)$ instead. As $A$ is assumed independent of $b$, the same must apply when $b \neq a$ as well. For the perfect correlations to hold, the similarly defined $B(b,\lambda)$ must share this dependence, $B(b,\lambda) = A(b,\lambda)$. Physically, this would imply that the information within $\lambda$ that determines the polarizations $A$ and $B$ is "duplicated" at the source and "carried" by each photon individually.

The second step of Bell's proof consists of writing the corresponding expression for the correlator,

$$(4) \qquad P_{\text{Bell}}(a,b) = \int d\lambda \rho(\lambda) A(a,\lambda) A(b,\lambda),$$

and deriving an inequality for correlators of this type. The derivation uses the facts that $\rho(\lambda)$ is never negative and is normalized, and that $A^2 = 1$. It involves introducing a third orientation $c$, and noting that from

$$(5) \qquad P_{\text{Bell}}(a,b) - P_{\text{Bell}}(a,c) = \int d\lambda \rho(\lambda) A(a,\lambda) A(b,\lambda)[1 - A(b,\lambda) A(c,\lambda)]$$

one may obtain, by taking the absolute value of the integrand,

$$(6) \qquad |P_{\text{Bell}}(a,b) - P_{\text{Bell}}(a,c)| \leq \int d\lambda \rho(\lambda)[1 - A(b,\lambda) A(c,\lambda)] = 1 - P_{\text{Bell}}(b,c).$$

This is the original Bell's inequality, adapted to photon polarizations.

The third step consists of substituting Eq. (2) for the correlators within the inequality, and noting that it is violated. Indeed, using nearby values of $b$ and $c$, the left hand side is linear in $|b-c|$ whereas the right hand side is quadratic, violating the inequality.

Thus, the quantum phenomena described by Eq. (2) are incompatible with any locally causal model, defying the expectations expressed by Einstein, in the EPR article and later.[40] Notice that the difficulty is not with the very use of a non-local description – such descriptions are common, e.g., the Liouville equation – but with the incompatibility with a more detailed locally causal description. Indeed, many phenomena which are typically considered peculiarly quantum, can be imitated by an underlying locally causal dynamics,[41] including the EPR phenomenon and quantum teleportation.[42] In contrast, the phenomena identified by Bell necessarily violate local causality.

III. Retro-causal model

Research activity on the topic of Bell's theorem was slow at first, but has intensified considerably in recent decades.[43] A "Bell inequality" involving four rather than three different orientations was derived from Eq. (4), i.e., for local deterministic hidden variables,[44] and then re-derived from Eq. (3), i.e., directly from local causality.[45] Violations of this inequality have been observed experimentally.[36] Imperfections in the experimental setup (called "loopholes") have been identified and addressed.[46] The QM formalism and its predictions have thus been eliminated from the discussion, yielding a direct demonstration that local causality (a.k.a. "local realism") is inconsistent with experimental observations.[17] Systems of three or four particles have been considered, and it was found[47] and experimentally demonstrated[48] that violations of local causality in some specific systems are even more explicit, i.e., can be identified without recourse to inequalities or to situations in which the QM predictions are probabilistic and cannot be made with 100% certainty. Related research has been performed in the fields of quantum computation and quantum cryptography.[49] Many articles are more philosophical, addressing, e.g., the nature of physical reality. Although Bell's mathematical approach can be used to illuminate such issues,[50] they can be very difficult to settle, absent significant technical advances, as exemplified by the controversy regarding the reality of atoms in the late 19$^{th}$ century.

Here we shall avoid all of these issues, and consider instead the following question: If locally causal descriptions are ruled out, what type of mathematical descriptions *can* reproduce QM? The theoretical description which served to motivate Bell, and was forcefully advocated by him later,[51] is Bohmian mechanics,[5] a redevelopment of an early idea of Louis de Broglie[52] which treats the wavefunction as a guiding wave for particle configurations. For $N$ particles, this description involves supplementing the wavefunction $\Psi(\mathbf{r}_1,\ldots,\mathbf{r}_N,t)$ by $N$ coordinates $\mathbf{R}_j(t)$. These configuration coordinates are chosen at random according to the standard $\Psi^+\Psi$ weights at the initial time (the notation allows for spin and spinors; else $|\Psi|^2$ would have been used), and then evolved according to the standard velocities, $(\hbar/M_j)\,\mathrm{Im}(\Psi^+\nabla_{\mathbf{r}_j}\Psi)/\Psi^+\Psi$, evaluated at the current configuration $\{\mathbf{r}_k = \mathbf{R}_k(t)\}_{k=1,\ldots,N}$ (here $M_j$ is the mass of the $j$th particle). The continuity equation associated with the Schroedinger evolution of $\Psi$ guarantees that the coordinates $\mathbf{R}_j(t)$ will continue to be distributed according to $\Psi^+\Psi$ at later times. Taking the $\mathbf{R}_j(t)$ as the predictions for positions at the time of measurement thus reproduces standard QM, as all quantum measurements can be reduced to measurements of positions.

Bohmian mechanics indeed has distinct advantages. Significantly, it does not suffer from the measurement problem. It also follows standard QM quite closely – the added coordinates are local quantities obeying local rules, and the non-local effects present are clearly due to the use of the wavefunction. In order to illustrate such non-local effects, Bell included in Ref. 13 a simplistic toy-model, which when adapted to photons, takes the initial condition $\lambda$ to be a uniformly distributed random angle, $\rho(\lambda) = const.$ and takes $B(b,\lambda)$ to be 1 if $|\lambda - b| < \pi/4$ and $-1$ otherwise (recall that the angles are defined modulo $\pi$). Nonlocality enters in the $b$-dependence of $A(a,b,\lambda)$, which is taken as 1 if $|\lambda - a'| < \pi/4$ and $-1$ otherwise, where the angle $a'$ is (somewhat artificially) chosen according to $1 - (4/\pi)(a - a') = \cos(2a - 2b)$. This model succeeds in reproducing the QM predictions of Eqs. (1) and (2): as the distribution of $\lambda$ is uniform, the correlator depends linearly on $|a - a'|$ in the range $[0, \pi/2]$. This is done by directly breaking locality, and without violating causality.

The realization that Bell's theorem assumes local causality, rather than merely locality, leads naturally to a consideration of mathematical descriptions which break causality altogether. Indeed, a similarly simplistic non-causal toy-model may be obtained[53] by taking $\lambda$ to be an angle which accepts one of the values $a$, $a + \pi/2$, $b$, $b + \pi/2$, with equal probabilities, i.e., by using

(7) $$\rho(\lambda|a,b) = \frac{1}{4}\left[\delta(\lambda - a) + \delta\left(\lambda - a - \frac{\pi}{2}\right) + \delta(\lambda - b) + \delta\left(\lambda - b - \frac{\pi}{2}\right)\right].$$

It is appropriate to take $\lambda$ to represent the linear polarization of the photons belonging to each pair. The model thus assumes that the photons are emitted by the source with polarizations which "anticipate" the directions of the apparatuses to be encountered in the future – a blatant and explicit violation of causality. The subsequent interaction with each apparatus follows the standard probability rules for single-photon polarization measurements (Malus' law):

(8) $$p(A|a,\lambda) = \begin{cases} \cos^2(a - \lambda) & A = 1 \\ \sin^2(a - \lambda) & A = -1 \end{cases},$$

together with the corresponding expression for $p(B|b,\lambda)$. It is straightforward to derive Eqs. (1) and (2) from Eqs. (7) and (8). In fact, each one of the possible values of $\lambda$ separately leads to the quantum-mechanical correlations of Eq. (2). For example, if $\lambda = a + \pi/2$ (the second possibility in Eq. (7)) then $A = -1$ with certainty, and $B = 1$ or $-1$ with probabilities of $\sin^2(a-b)$ and $\cos^2(a-b)$ respectively (when $b = a$ or $b = a + \pi/2$, the values of both $A$ and $B$ are selected with certainty, resulting in perfect correlations). The resulting contribution to the correlator, $\langle AB \rangle_{\lambda=a+\pi/2} = -\sin^2(a-b) + \cos^2(a-b)$, is equal to $P_{QM}(a,b)$. As this result is obtained for each one of the four possible values of $\lambda$, any model which makes these choices with other weights, including models which would choose one of the possible values of $\lambda$ with probability 100%, would reproduce Eq. (2). It is necessary to introduce $\lambda$ as a *stochastic* variable with equal probabilities for $\lambda = a$ and for $\lambda = a + \pi/2$, and similarly for $\lambda = b, b + \pi/2$, in order to reproduce also Eq. (1), e.g., through $\frac{1}{2}\langle B \rangle_{\lambda=a} + \frac{1}{2}\langle B \rangle_{\lambda=a+\pi/2} = \frac{1}{2}(\cos^2(a-b) - \sin^2(a-b)) + \frac{1}{2}(\sin^2(a-b) - \cos^2(a-b)) = 0$. In Eq. (7), equal weights have been chosen for *all* values of $\lambda$, for symmetry reasons (aesthetics). If one of the measurements, say the one on the left, is performed before the other, then avoiding causal loops requires[54] using an asymmetric version of the toy-model, with $\rho(\lambda|a) = \frac{1}{2}\left[\delta(\lambda - a) + \delta\left(\lambda - a - \frac{\pi}{2}\right)\right]$ instead of Eq. (7) (i.e., $\lambda$ is independent of $b$). No change in the prescription for the probabilities of $A$ and $B$, Eq. (8), is required.

This retro-causal model is appealing in its simplicity. As should be expected for such models,[34] $\lambda$ represents an inaccessible quantity, which can not be simply measured at the source – if it were measurable, one could send signals to $a$ and $b$ instructing them to avoid the values $\lambda, \lambda + \pi/2$, resulting in a paradoxical situation. For a source consisting of a Ca atom, one may easily imagine an attempt to measure the polarization $\lambda$ by interrogating the intermediate $J = 1$ state of the radiative cascade, and measuring the relevant component of its spin. This would destroy the EPR correlations of the photon pair, just as a "which path" measurement destroys the interference pattern in a two-slit experiment. The inaccessibility of $\lambda$ thus reflects that of the "which path" variables in standard QM, and should not be considered surprising.

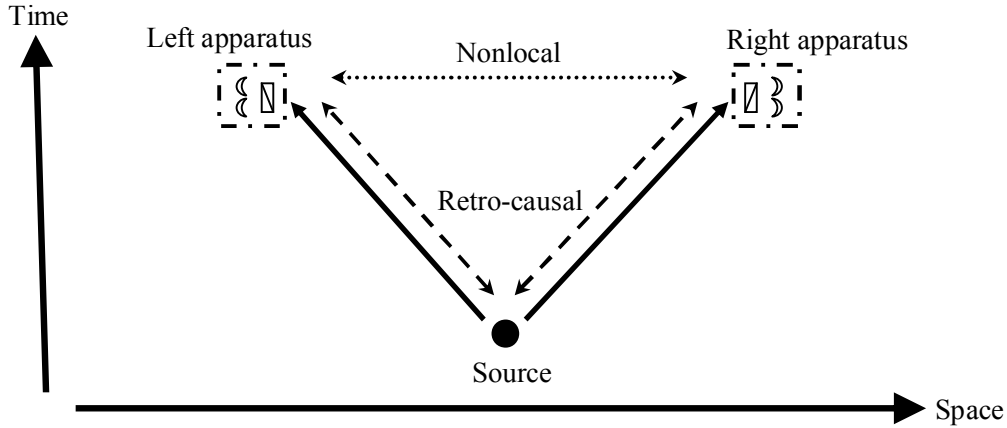

Fig 2: Space-time sketch comparing alternatives for mathematical models of quantum correlations. The full arrows describe the paths of the photons from the source to the measuring apparatuses. In directly nonlocal models (dotted double arrow) the mathematical description of the state of the system propagates linearly from the past into the future, but output variables corresponding (e.g.) to the left apparatus may depend on variables describing the right apparatus. In retro-causal models (dashed arrows) propagation of mathematical information from the apparatuses backwards in time to the source is allowed, and thus no instantaneous "action at a distance" is needed. In these models, the variables which carry information which has propagated into the past represent microscopic physical quantities (e.g., attributes of "quantum fluctuations"), which must be inaccessible to the macroscopic level – otherwise, causal paradoxes would arise.

Note that the assumptions of locality and causality are inseparable in two distinct senses (see Fig. 2). Firstly, they imply each other, in the sense that (a) violations of causality imply indirect violations of locality (if $\lambda$ depends on $a$ and $B$ depends on $\lambda$, then $B$ depends indirectly on $a$), and (b) in a relativistic setting, violations of locality which are instantaneous in one reference frame violate causality in other reference frames. Secondly, within the framework considered, any non-causal model can be "translated" into a non-local model, and vice versa. For example, $\lambda$ of Eq. (7) can be replaced by a random integer $n$ from 1 to 4, thus becoming independent of $a$ and $b$, and then an angle $\lambda'(n,a,b)$ can be defined as associated with a later time, but in a manner such that $\lambda'$ plays precisely the role of the previous $\lambda$. Conversely, any non-local model can be "translated" into a retro-causal one, by evaluating $A$ and $B$ "at the source," and including them as elements of the set $\lambda$.[55] Although the class of non-local models and the class of non-causal models are thus "mathematically equivalent," physically, locality and causality are quite distinct. In particular, direct non-locality clashes with the principles of relativity.

|  | **"Initial conditions"** $\lambda$ | **Dynamics determining results** $A$ **and** $B$ |
|---|---|---|
| **Quantum Mechanics** | $\lambda$ is the wavefunction $\psi$. | **Stochastic** and **non-local**, $P(A,B|a,b,\lambda)$ provided. |
| **Bohmian Mechanics** | $\lambda$ consists of the wavefunction $\psi$ and initial particle positions $\mathbf{R}_i$. The coordinates $\mathbf{R}_i$ are **stochastic**, distributed according to $\psi$. | **Deterministic** and **non-local**, $A(a,b,\lambda)$ and $B(a,b,\lambda)$ specified by the dynamics. |
| **Bell's non-local toy-model** | $\lambda \in [0,\pi)$ is an angle, and is **stochastic**, uniformly distributed. | **Deterministic** and **non-local**, $A(a,b,\lambda)$ and $B(b,\lambda)$ provided. |
| **Retro-causal toy-model** | $\lambda \in [0,\pi)$ is an angle. It depends on $a$ and $b$ in a **stochastic**, **retro-causal** manner. $P(\lambda|a,b)$ is discrete, Eq. (7). | **Stochastic** (except for $a = \lambda, \lambda + \pi/2$), $P(A|a,\lambda)$ and $P(B|b,\lambda)$ provided by Eq. (8). |

Table 1: Different ways of reproducing the Bell's-inequality-violating statistics of Eqs. (1) and (2). Bell's non-local model is naturally generalized by Bohmian mechanics. Finding general non-causal models remains a challenge.

IV.  Concluding remarks

We have seen that mathematical models which reproduce the quantum correlations of pairs of photons [Eqs. (1) and (2)] can be either directly non-local or retro-causal. A more detailed comparison is given in Table 1, which compares standard QM, Bohmian mechanics,[56] Bell's non-local toy-model, and the retro-causal toy-model. The first three are causal and directly non-local descriptions. They differ in the manner in which probabilities enter: in QM, the results are probabilistic, whereas in Bohmian mechanics and in the non-local toy-model the initial conditions are random and the dynamic rules are deterministic. The last line corresponds to the new non-causal toy-model [Eqs. (7) and (8)]. Additional models could be included in the table. For example, in the 1950s, Bohm considered also a description in which the Schroedinger wavefunction serves to guide particles with diffusive, rather than deterministic, dynamics.[57] It would be especially relevant to include additional retro-causal descriptions, but those available in the literature[30,58,59] are not valid candidates for the present comparison, as they do not provide explicit expressions for the measurement results *A* and *B* or their probabilities, treating them instead as given "final conditions."

The comparison of Table 1 leads naturally to the question: Can the retro-causal, stochastic toy-model be generalized to all quantum phenomena, in analogy with the fact that Bohmian mechanics provides a general description belonging to the same "deterministic-nonlocal" class as the simple model of Bell? While a direct generalization would be desirable, it would not resolve the measurement problem, as the variables associated with measurement apparatuses are treated here quite differently from other physical variables. A more ambitious goal would be to seek a reformulation which, at a fundamental level, describes quantum fluctuations which are not subject to the causal arrow of time. Achieving agreement with the predictions of QM, with the macroscopic causal arrow of time and with the rule of no non-local signaling,[60] would require breaking the time-reversal symmetry, perhaps by postulating a low entropy in the past and including dissipation in the description of measurement devices, in analogy to current descriptions of heat baths.[61]

When compared to standard QM, these ideas lead to the view that QM is a description following only the irreversibly registered part of the information regarding a system. Representing all of this information up to a given instant in time requires the exponentially complex structure of the wavefunction.[62] Such a view makes it appear natural to have conservation of information – unitary evolution – punctuated by "jumps" whenever additional information is registered. In contrast, the retro-causal descriptions of general quantum phenomena available in the literature,[10,30] employ the wavefunctions in a manner which allows them to propagate backwards in time. When applied to systems with several particles, such descriptions would involve a preferred reference frame, and thus would not avoid the clash with relativity. Moreover, use of wavefunctions involves direct non-locality, instead of taking advantage of the indirect route offered by retro-causality. It is clearly desirable to go beyond this.

These concluding remarks are very much in line with Bell's own later views, as evidenced by the concluding paragraph of his 1990 recapitulation:[23]

> "The more closely one looks at the fundamental laws of physics, the less one sees of the laws of thermodynamics. The increase of entropy emerges only for large complicated systems, in an approximation depending on 'largeness' and 'complexity.' Could it be that causal structure emerges only in something like a 'thermodynamic' approximation, where the notions 'measurement' and 'external field' become

> legitimate approximations? Maybe that is part of the story, but I do not think it can be all. Local commutativity does not for me have a thermodynamic air about it. …"

It is argued here that as Bell took the causal arrow of time for granted, he over-interpreted local commutativity, as indeed this time-reversal symmetric property is routinely taken to represent local causality, which is not symmetric. Adding retro-causation to the picture, possibly together with stochasticity, may perhaps provide "the other part of the story." To quote Bell again,[63]

> "Let us hope that these analyses may one day be illuminated, perhaps harshly, by some simple constructive model. … what is proved by impossibility proofs is lack of imagination."

Given the persistence of pre-20th-century ideas on causality and determinism, it is likely that many readers of this article will not entertain high hopes for this line of enquiry. However, even the unconvinced should heed the call to prominently mention the symmetry-breaking causal arrow of time whenever the assumptions leading to Bell's theorem are discussed.

*Acknowledgements*

The author thanks D. Aharonov, J. Berkovitz, A.C. Elitzur, O.J.E. Maroney, T. Norsen, E. Pazy, H. Price, R. Sutherland, H. Westman, K. Wharton and A. Zeilinger for helpful discussions.